# Linear-algebraic list decoding of folded Reed-Solomon codes

VENKATESAN GURUSWAMI[*]

Computer Science Department
Carnegie Mellon University
Pittsburgh, PA 15213

**Abstract**

Folded Reed-Solomon codes are an explicit family of codes that achieve the optimal trade-off between rate and error-correction capability: specifically, for any $\varepsilon > 0$, the author and Rudra (2006,08) presented an $n^{O(1/\varepsilon)}$ time algorithm to list decode appropriate folded RS codes of rate $R$ from a fraction $1 - R - \varepsilon$ of errors. The algorithm is based on multivariate polynomial interpolation and root-finding over extension fields. It was noted by Vadhan that interpolating a linear polynomial suffices if one settles for a smaller decoding radius (but still enough for a statement of the above form). Here we give a simple linear-algebra based analysis of this variant that eliminates the need for the computationally expensive root-finding step over extension fields (and indeed any mention of extension fields). The entire list decoding algorithm is linear-algebraic, solving one linear system for the interpolation step, and another linear system to find a small subspace of candidate solutions. Except for the step of pruning this subspace, the algorithm can be implemented to run in *quadratic* time.

The theoretical drawback of folded RS codes are that both the decoding complexity and proven worst-case list-size bound are $n^{\Omega(1/\varepsilon)}$. By combining the above idea with a pseudorandom subset of all polynomials as messages, we get a Monte Carlo construction achieving a list size bound of $O(1/\varepsilon^2)$ which is quite close to the existential $O(1/\varepsilon)$ bound (however, the decoding complexity remains $n^{\Omega(1/\varepsilon)}$).

Our work highlights that constructing an explicit *subspace-evasive* subset that has small intersection with low-dimensional subspaces — an interesting problem in pseudorandomness in its own right — could lead to explicit codes with better list-decoding guarantees.

## 1 Introduction

Reed-Solomon (RS) codes are an important family of error-correcting codes with many applications in theory and practice. An $[n, k]_q$ RS code over the field $\mathbb{F}_q$ with $q$ elements encodes polynomials $f \in \mathbb{F}_q[X]$ of degree at most $k - 1$ by its evaluations at $n$ distinct elements from $\mathbb{F}_q$. The encodings of any two distinct polynomials differ on at least $n - k + 1$ positions, which bestows the

[*]Research supported in part by a Packard Fellowship and NSF grants CCF-0953155 and CCF-0963975. Email: guruswami@cmu.edu. Any opinions, findings, and conclusions or recommendations expressed in this material are those of the author(s) and do not necessarily reflect the views of the National Science Foundation.

RS code with an error-correction capability of $(n-k)/2$ worst-case errors. Classical algorithms, the first one due to Peterson [20] over 50 years ago, are able to decode such a RS code from up to $(n-k)/2$ errors (i.e., a fraction $(1-R)/2$ of errors where $R = k/n$ is the rate code) in polynomial time.

Decoding beyond the radius $(1-R)/2$ is not possible if the decoder is required to always identify the correct message unambiguously. However, allowing the decoder to output a small list in the worst-case enables decoding well beyond this bound. This notion is called list decoding, and has been an actively researched topic in the last decade. It has found many applications in complexity theory and pseudorandomness (see [23, 24, 26] for some surveys) beyond its direct relevance to error-correction and communication.

For RS codes, Sudan [22] gave a list decoding algorithm to decode beyond the $(1-R)/2$ radius for rates $R < 1/3$. For rates $R \to 0$, the algorithm could correct a fraction of errors approaching $1$, a remarkable feature that led to many complexity-theoretic applications. The author and Sudan [14] improved the error-correction radius to $1-\sqrt{R}$, matching the so-called "Johnson radius," which is the a priori lower bound on list-decoding radius of a code as a function of its distance alone. This result improved upon the traditional $(1-R)/2$ bound for *all* rates. The $1-\sqrt{R}$ bound remains the best error-correction radius achievable to date for list decoding RS codes.

A standard random coding argument, however, shows the *existence* of rate $R$ codes $C \subseteq \Sigma^n$ list-decodable even up to radius $1-R-\varepsilon$. Specifically, $C$ has the combinatorial property that for every $y \in \Sigma^n$, there are at most $L = O(1/\varepsilon)$ codewords of $C$ within Hamming distance $(1-R-\varepsilon)n$ from $y$. Here $\varepsilon > 0$ can be an arbitrarily small constant. The quantity $L$ is referred to as the *list-size*. Note that $1-R$ is a clear information-theoretic limit for error-correction, since at least $\approx Rn$ received symbols must be correct to have any hope of recovering the $Rn$ message symbols.

A few years back the author and Rudra, building upon the work of Parvaresh and Vardy [19], gave an *explicit* construction of codes of rate $R$ which are list-decodable in polynomial time up to radius $1-R-\varepsilon$, with a list-size of $n^{O(1/\varepsilon)}$ [13]. These codes were a "folded" version of Reed-Solomon codes, defined below.

**Definition 1** (*m*-folded Reed-Solomon code). *Let $\gamma \in \mathbb{F}_q$ be a primitive element of $\mathbb{F}_q$. Let $n \leqslant q-1$ be a multiple of $m$, and let $1 \leqslant k < n$ be the* degree parameter.

*The folded Reed-Solomon (FRS) code $\mathrm{FRS}_q^{(m)}[n,k]$ is a code over alphabet $\mathbb{F}_q^m$ that encodes a polynomial $f \in \mathbb{F}_q[X]$ of degree $k-1$ as[1]*

$$f(X) \mapsto \left( \begin{bmatrix} f(1) \\ f(\gamma) \\ \vdots \\ f(\gamma^{m-1}) \end{bmatrix}, \begin{bmatrix} f(\gamma^m) \\ f(\gamma^{m+1}) \\ \vdots \\ f(\gamma^{2m-1}) \end{bmatrix}, \ldots, \begin{bmatrix} f(\gamma^{n-m}) \\ f(\gamma^{n-m+1}) \\ \vdots \\ f(\gamma^{n-1}) \end{bmatrix} \right). \qquad (1)$$

Observe that the FRS code has block length $N = n/m$ and rate $R = k/n$ (equal to the rate of the original, unfolded Reed-Solomon code, which corresponds to the choice $m = 1$). For any integer $s$, $1 \leqslant s \leqslant m$, a list decoding algorithm for the above FRS codes for a fraction $\approx 1 - \left(\frac{mR}{m-s+1}\right)^{s/(s+1)}$ of errors is presented in [13], with decoding complexity $q^{O(s)}$ and list-size $q^s$. The result of [13] can also be viewed as a better algorithm for decoding Reed-Solomon codes when the errors occur in bursts, since the evaluation points of the RS encoding are usually ordered as powers of $\gamma$ for some primitive $\gamma$.

---

[1] The actual code depends also on the choice of the primitive element $\gamma$. But the results hold for any choice of primitive $\gamma$, so for notational convenience we suppress the dependence on $\gamma$ and assume some canonical choice of $\gamma$ is fixed.

For suitably large constants $s, m$ depending on $\varepsilon$, the above list decoding radius for FRS codes exceeds $1 - R - \varepsilon$. However, the list-size bound then becomes $n^{\Omega(1/\varepsilon)}$ which has a rather poor dependence on the distance $\varepsilon$ to the optimal trade-off. Improving the list-size is therefore an important open question. Recall that existentially a list-size as small as $O(1/\varepsilon)$ is possible. The decoding algorithms in [19, 13] consist of two steps (see Section 2.1 for more details): (i) multivariate polynomial interpolation (to find an algebraic equation that candidate message polynomials $f$ must satisfy), and (ii) solving this equation via root-finding over extension fields. The interpolation step reduces to finding a nonzero solution to a homogeneous linear system, and theoretically the second step is the computationally more expensive one.

Vadhan showed recently that a weaker decoding radius (which however still suffices to list decode up to radius $1 - R - \varepsilon$) can be achieved by a simplified interpolation step that only interpolates a degree 1 multivariate polynomial [25]. Further, there is *no need* to use multiplicities in the interpolation as in the earlier algorithms [14, 19, 13].[2] This offers a clean and simple exposition of a list decoding algorithm for FRS codes (that can be viewed as a multidimensional version of the Welch-Berlekamp decoder for RS codes) for a fraction $\approx \frac{s}{s+1}\left(1 - \frac{mR}{m-s+1}\right)$ of errors [10] (see Section 2.2). The second root-finding step of the decoder, however, remained unchanged.

**Contributions of this work.** Here, we note that this Welch-Berlekamp style "degree 1" list decoder, not only offers a simpler exposition, but also offers some promising advantages. Our starting point is the simple observation that in this case the candidate solutions to the algebraic equation form an affine subspace (of the full message space $\mathbb{F}_q^k$). This implies that the second step of the list decoding can also be tackled by solving a linear system!

By inspecting the structure of this linear system, we give an elementary linear-algebraic proof (Lemma 6) that the subspace of solutions has dimension at most $s-1$, a fact that was earlier proved by root counting over extension fields in [13, 25]. This shows that the exponential dependence in $s$ of the list-size bound was inherently because of the dimension of the interpolation (and it wasn't crucial that we had the identity $f(\gamma^{s-1} X) = f(X)^{q^{s-1}}$ over some extension field[3]).

The linear-algebraic proof also gives a *quadratic* time algorithm to find a basis for the subspace (instead of the cubic time of Gaussian elimination). This leads to a quadratic runtime for the list decoder, *except* for the final step of pruning the subspace to actually find the close-by codewords (formal statement in Theorem 7). This pruning step needs to check each element of the subspace and thus unfortunately could still take $q^s$ time. However, in practice (or when errors occur randomly), the dimension of the output subspace will likely be very small, probably even 0 (implying a unique solution), and in such cases we get significant gains in efficiency compared to [13].

---

[2]However, the method of multiplicities is still crucial if one wants a soft-decision list decoder, which, at least for Reed-Solomon codes, has been a very influential development [17], with many subsequent papers looking at practical decoding architectures.

[3]This identity, however, seems to be the only known way to bound the list-size when higher degrees are used in the interpolation.

**Better list-size via subspace-evasive sets.** Our second contribution is to exploit the subspace structure of the candidate solutions to improve the list-size bound. The idea is to restrict the coefficient vectors of the message polynomial to a large "subspace-evasive" subset that has small intersection with subspaces of low dimension. Subspace-evasive sets seem like fundamental combinatorial objects interesting in their own right. They are related to affine extractors, and also have applications to constructing bipartite Ramsey graphs [21]. As one would expect, a random set has excellent subspace-evasiveness, but finding good explicit constructions is wide open. Our application to list decoding in this work provides another motivation for the interesting problem of constructing subspace-evasive sets.

Using a pseudorandom construction of subspace-evasive subsets (in fact, algebraic varieties) based on limited independence, we give a Monte Carlo construction (succeeding with high probability) of rate $R$ codes list-decodable up to a fraction $1 - R - \varepsilon$ of errors with a list-size of $O(1/\varepsilon^2)$ (Theorem 10 gives the exact statement). Due to the pruning step, the worst-case runtime is however still $n^{\Omega(1/\varepsilon)}$. Nevertheless, this is the first construction with a better than $n^{\Omega(1/\varepsilon)}$ list-size for decoding up to the information-theoretic limit of $1 - R - \varepsilon$ fraction of errors.

For this construction, we do not know a polynomial time computable encoding function that maps messages to polynomials in the subspace-evasive subset. However, if we settle for a list-size of $O(n)$ — still much better than the earlier $n^{\Omega(1/\varepsilon)}$ bound — a polynomial time encoder can also be obtained. We stress that only our code construction is randomized, and once it succeeds (which happens w.h.p.), the list decoding properties hold for every received word and the encoding and list decoding procedures run in deterministic polynomial time.

**Organization.** We describe the list decoding algorithm for FRS codes and our linear-algebraic analysis of it in Section 2. We make some related remarks about the linear algebra approach in Section 3. We use subspace-evasive sets to give our Monte Carlo construction of codes achieving list decoding capacity with improved list-size in Section 4.

## 2 List decoding folded Reed-Solomon codes

Suppose a codeword of the $m$-folded RS code (Definition 1) was transmitted and we received a string in $\mathbf{y} \in (\mathbb{F}_q^m)^N$ which we view as an $m \times N$ matrix over $\mathbb{F}_q$:

$$\begin{pmatrix} y_0 & y_m & & y_{n-m+1} \\ y_1 & y_{m+1} & & \vdots \\ y_2 & y_{m+2} & & \vdots \\ & & \ddots & \\ y_{m-1} & \cdots & & y_{n-1} \end{pmatrix} \qquad (2)$$

We would like to recover a list of all polynomials $f \in \mathbb{F}_q[X]$ of degree $k - 1$ whose folded RS encoding (1) agrees with $\mathbf{y}$ in at least $N - e$ columns, for some error bound $e$. Note that an agreement means that *all* $m$ values in that particular column match. The following theorem is from [13].

**Theorem 2.** *For every integer $s$, $1 \leqslant s \leqslant m$ and any constant $\delta > 0$, there is a list decoding algorithm for*

the folded Reed-Solomon code $\text{FRS}_q^{(m)}[n,k]$ that list decodes from up to $e$ errors as long as

$$e \leqslant N - (1+\delta)\frac{(k^s N(m-s+1))^{1/(s+1)}}{m-s+1}$$

where $N = n/m$ is the block length of the code. The algorithms runs in $(O_\delta(q))^{O(s)}$ time and outputs a list of size at most $q^s$.

Note that the fraction of errors corrected by this algorithm as a function of the rate $R = k/n = k/(Nm)$ is

$$1 - (1+\delta)\left(\frac{mR}{m-s+1}\right)^{s/(s+1)}. \tag{3}$$

By picking $\delta \approx \varepsilon$, $s \approx 1/\varepsilon$ and $m \approx 1/\varepsilon^2$, the above quantity is at least $1 - R - \varepsilon$, and the decoding complexity and list size are $\approx q^{O(1/\varepsilon)}$.

## 2.1 Overview of above decoding algorithm

We briefly recap the high level structure of this decoding algorithm. The quantity $s$ is a parameter of the algorithm. In the first step, the algorithm interpolates a multivariate polynomial $Q \in \mathbb{F}_q[X, Y_1, Y_2, \ldots, Y_s]$ of low *weighted degree* (where the $Y_i$'s have weight $k-1$ and $X$ has weight 1) such that, for every $i$, $0 \leqslant i \leqslant n-s$, $Q(X, Y_1, \ldots, Y_s)$ vanishes at $(\gamma^i, y_i, y_{i+1}, \ldots, y_{i+s-1}) \in \mathbb{F}_q^{s+1}$ with high multiplicity (related to the other parameter $\delta$ of the algorithm). This step can be accomplished by solving a homogeneous linear system over $\mathbb{F}_q$. The degree and multiplicity parameters in the interpolation step are carefully picked to ensure the following two properties: (i) a nonzero $Q$ meeting the interpolation requirements exists, and (ii) every $f \in \mathbb{F}_q[X]$ of degree at most $(k-1)$ whose FRS encoding agrees with $\mathbf{y}$ on at least $N - e$ places (and which, therefore, must be output by the list decoder) satisfies the functional equation

$$Q(X, f(X), f(\gamma X), \cdots, f(\gamma^{s-1} X)) = 0.$$

In the second step of the decoder, all solutions $f$ to the above equation are found. This is done by observing that $f(\gamma X) = f(X)^q \pmod{E(X)}$ where $E(X) = (X^{q-1} - \gamma)$, and therefore $f \mod E(X)$ can be found by finding the roots of the univariate polynomial

$$T(Y) = Q(X, Y, Y^q, \ldots, Y^{q^{s-1}}) \mod E(X)$$

with coefficients from $L = \mathbb{F}_q[X]/(E(X))$. The polynomial $E(X)$ is irreducible over $\mathbb{F}_q$ and therefore $L$ is an *extension field*. The parameter choices ensure that $T \neq 0$, and thus $T$ cannot have too many roots and these roots may all be found in polynomial time. Finally, this list is pruned to only output those polynomials whose FRS encoding is in fact close to the received word $\mathbf{y}$.

## 2.2 A Welch-Berlekamp style interpolation

We will now describe a variant of the above scheme where the interpolation step will fit a non-zero "linear" polynomial $Q(X, Y_1, Y_2, \ldots, Y_s)$ (with degree 1 in the $Y_i$'s). This can be viewed as a higher-dimensional generalization of the Welch-Berlekamp algorithm [27, 8]. This elegant version is due to Vadhan and is described in his monograph [25, Chap. 5] (and also used in the author's lecture notes [10]). For completeness, and because it will be convenient to refer to it in the second step, we give a self-contained presentation here.

The original motivation for this variant was that it had simpler parameter choices and an easier exposition (even though the error-correction guarantee worsened, it still allowed approaching a decoding radius of $1 - R$ in the limit). In particular, it has the advantage of *not* requiring the use of multiplicities in the interpolation. (Essentially, the freedom to do $s$-variate interpolation for a parameter $s$ of our choosing allows us to work with simple interpolation while still gaining in error-correction radius with increasing $s$. This phenomenon also occurred in one of the algorithms in [12] for list decoding correlated algebraic-geometric codes.)

In this work, our contribution is to put the simple linear structure of the interpolated polynomial to good use and exploit it to substitute the root-finding step with a more efficient step of solving a linear system.

Given a received word as in (2) we will interpolate a nonzero polynomial

$$Q(X, Y_1, Y_2, \ldots, Y_s) = A_0(X) + A_1(X)Y_1 + A_2(X)Y_2 + \cdots + A_s(X)Y_s \tag{4}$$

over $\mathbb{F}_q$ with the degree restrictions $\deg(A_i) \leqslant D$ for $i = 1, 2, \ldots, s$ and $\deg(A_0) \leqslant D + k - 1$, where the degree parameter $D$ is chosen to be

$$D = \left\lfloor \frac{N(m - s + 1) - k + 1}{s + 1} \right\rfloor . \tag{5}$$

The number of monomials in a polynomial $Q$ with these degree restrictions equals

$$(D + 1)s + D + k = (D + 1)(s + 1) + k - 1 > N(m - s + 1) \tag{6}$$

for the above choice (5) of $D$. The interpolation requirements on $Q \in \mathbb{F}_q[X, Y_1, \ldots, Y_s]$ are the following:

$$Q(\gamma^{im+j}, y_{im+j}, y_{im+j+1}, \cdots, y_{im+j+s-1}) = 0 \quad \text{for } i = 0, 1, \ldots, n/m - 1, \ j = 0, 1, \ldots, m - s . \tag{7}$$

Since the number of interpolation conditions $(n/m) \cdot (m - s + 1)$ is less than the number of degrees of freedom (monomials) in $Q$, we can conclude the following. The claim about the near-linear runtime has been shown in [4] (see Proposition 5.11 in Chapter 5).

**Lemma 3.** *A nonzero $Q \in \mathbb{F}_q[X, Y_1, \ldots, Y_s]$ of the form (4) satisfying the interpolation conditions (7) can be found by solving a homogeneous linear system over $\mathbb{F}_q$ with at most $Nm$ constraints and variables. Further this interpolation can be performed in $O(Nm \log^2(Nm) \log \log(Nm))$ operations over $\mathbb{F}_q$.*

The following lemma shows that any such polynomial $Q$ gives an algebraic condition that the message polynomials $f(X)$ we are interested in list decoding must satisfy.

**Lemma 4.** *If $f \in \mathbb{F}[X]$ is a polynomial of degree at most $k - 1$ whose FRS encoding (1) agrees with the received word $\mathbf{y}$ in at least $t$ columns for $t > \frac{D+k-1}{m-s+1}$, then*

$$Q(X, f(X), f(\gamma X), \ldots, f(\gamma^{s-1}X)) = 0 . \tag{8}$$

*Proof.* Define $R(X) = Q(X, f(X), f(\gamma X), \ldots, f(\gamma^{s-1}X))$. Due to the degree restrictions on $Q$, the degree of $R(X)$ is easily seen to be at most $D + k - 1$. If the FRS encoding of $f$ agrees with $\mathbf{y}$ in the $i$'th column (for some $i \in \{0, 1, \ldots, N-1\}$), we have

$$f(\gamma^{im}) = y_{im}, \quad f(\gamma^{im+1}) = y_{im+1}, \quad \cdots \quad , f(\gamma^{im+m-1}) = y_{im+m-1} \ .$$

Together with the interpolation conditions (7), this implies $R(\gamma^{im+j}) = 0$ for $j = 0, 1, \ldots, m-s$. In other words $R$ picks up at least $m - s + 1$ distinct roots for each such column $i$. Thus $R$ must have at least $t(m - s + 1)$ roots in all. Since $\deg(R) \leqslant D + k - 1$, if $t > (D + k - 1)/(m - s + 1)$, we must have $R = 0$. □

For the choice of $D$ in (5), the requirement on $t$ in Lemma 4 is met if $t(m-s+1) > \frac{N(m-s+1)+s(k-1)}{s+1}$, and hence if

$$t \geqslant \frac{N}{s+1} + \frac{s}{s+1}\frac{k}{m-s+1} = N\left(\frac{1}{s+1} + \frac{s}{s+1}\frac{mR}{m-s+1}\right) \ . \tag{9}$$

In other words, the fractional agreement needed is $\frac{1}{s+1} + \frac{s}{s+1}\frac{mR}{m-s+1}$. Note that by the AM-GM inequality, this agreement is always higher than the agreement fraction $\left(\frac{mR}{m-s+1}\right)^{s/(s+1)}$ needed in (3).[4] Thus this variant corrects a smaller fraction of errors. Nevertheless, with the choice $s \approx 1/\varepsilon$ and $m \approx 1/\varepsilon^2$, the fraction of errors corrected can still exceed $1 - R - \varepsilon$. Further, as we see next, it offers some advantages when it comes to retrieving the solutions $f$ to (8).

### 2.3 Retrieving candidate polynomials $f$

By the preceding section, to complete the list decoding we need to find all polynomials $f \in \mathbb{F}_q[X]$ of degree at most $k - 1$ that satisfy

$$A_0(X) + A_1(X)f(X) + A_2(X)f(\gamma X) + \cdots + A_s(X)f(\gamma^{s-1}X) = 0 \ . \tag{10}$$

We note the following simple but very useful fact:

**Observation 5.** *The above is a system of linear equations over $\mathbb{F}_q$ in the coefficients $f_0, f_1, \cdots, f_{k-1}$ of the polynomial $f(X) = f_0 + f_1 X + \cdots + f_{k-1} X^{k-1}$. Thus, the solutions $(f_0, f_1, \ldots, f_{k-1})$ of (10) form an affine subspace of $\mathbb{F}_q^k$.*

In particular, the above immediately gives an efficient algorithm to find a compact representation of all the solutions to (10) — simply solve the linear system! This simple observation is the starting point driving this work.

We next prove that when $\gamma$ is primitive, the space of solutions has dimension at most $s - 1$. Note that we *already* knew this by the earlier argument over the extension field $\mathbb{F}_q[X]/(X^{q-1} - \gamma)$. But it is instructive to give a direct proof of this working only over $\mathbb{F}_q$. The proof in fact works when the order of $\gamma$ is at least $k$. Further, it exposes the simple structure of the linear system which can be used to find a basis for the solutions in quadratic time.

---

[4]Recall that for Reed-Solomon codes ($m = 1$) this was also exactly the case: the classical algorithms unique decoded the codeword when the agreement fraction was at least $\frac{1+R}{2}$, and the list decoding algorithm in [14] list decoded from agreement fraction $\sqrt{R}$.

**Lemma 6.** *If the order of $\gamma$ is at least $k$ (in particular when $\gamma$ is primitive), the affine space of solutions to (10) has dimension $d$ at most $s - 1$.*

*Further, one can compute using $O((Nm)^2)$ field operations over $\mathbb{F}_q$ a matrix $M \in \mathbb{F}_q^{k \times d}$ (for some $d \leqslant s - 1$) and a vector $\mathbf{z} \in \mathbb{F}_q^k$ such that the solutions are contained in the affine space $M\mathbf{x} + \mathbf{z}$ for $\mathbf{x} \in \mathbb{F}_q^d$. Also, the matrix $M$ can be assumed to have the $d \times d$ identity matrix as a submatrix (without any extra computation).*

*Proof.* First, by factoring out a common powers of $X$ that divide all of $A_0(X), A_1(X), \ldots, A_s(X)$, we can assume that at least one $A_{i^*}(X)$ for some $i^* \in \{0, 1, \ldots, s\}$ is not divisible by $X$, and has nonzero constant term. Further, if $A_1(X), \ldots, A_s(X)$ are all divisible by $X$, then so is $A_0(X)$, so we can take $i^* > 0$.

Let us denote $A_i(X) = \sum_{j=0}^{D+k-1} a_{i,j} X^j$ for $0 \leqslant i \leqslant s$. (We know that the degree of $A_i(X)$ for $i \geqslant 1$ is at most $D$, so $a_{i,j} = 0$ when $i \geqslant 1$ and $j > D$, but for notational ease let us introduce these coefficients.) Define the polynomial

$$B(X) = a_{1,0} + a_{2,0}X + a_{3,0}X^2 + \cdots + a_{s,0}X^{s-1} \ .$$

We know that $a_{i^*,0} \neq 0$, and therefore $B \neq 0$.

We will prove our upper bound on the rank of the solution space by examining the condition that the coefficients of $X^r$ of the polynomial

$$\Lambda(X) = A_0(X) + A_1(X)f(X) + A_2(X)f(\gamma X) + \cdots + A_s(X)f(\gamma^{s-1}X)$$

on the left hand side of (10) equals $0$ for $r = 0, 1, 2, \ldots$.

The constant term of $\Lambda(X)$ equals $a_{0,0} + a_{1,0}f_0 + a_{2,0}f_0 + \cdots + a_{s,0}f_0 = a_{0,0} + B(1)f_0$. Thus if $B(1) \neq 0$, then $f_0$ is uniquely determined as $-a_{0,0}/B(1)$. If $B(1) = 0$, then $a_{0,0} = 0$ or else there will be no solutions to (10) and in that case $f_0$ can take an arbitrary value in $\mathbb{F}_q$.

The coefficient of $X^r$ of $\Lambda(X)$ equals

$$a_{0,r} + f_r \cdot (a_{1,0} + a_{2,0}\gamma^r + \cdots + a_{s,0}\gamma^{(s-1)r}) + f_{r-1} \cdot (a_{1,1} + a_{2,1}\gamma^{r-1} + \cdots + a_{s,1}\gamma^{(s-1)(r-1)}) + \quad (11)$$
$$\cdots + f_1 \cdot (a_{1,r-1} + a_{2,r-1}\gamma + \cdots + a_{s,r-1}\gamma^{s-1}) + f_0 \cdot (a_{1,r} + \cdots + a_{s,r})$$
$$= B(\gamma^r)f_r + \left(\sum_{i=0}^{r-1} b_i^{(r)} f_i\right) + a_{0,r} \quad (12)$$

for some coefficients $b_i^{(r)} \in \mathbb{F}_q$. The linear form (12) must thus equal $0$. The key point is that if $B(\gamma^r) \neq 0$, then this implies that $f_r$ is an affine combination of $f_0, f_1, \ldots, f_{r-1}$ and in particular is uniquely determined given values of $f_0, f_1, \ldots, f_{r-1}$.

Thus the dimension of the space of solutions is at most the number of $r$, $0 \leqslant r < k$, for which $B(\gamma^r) = 0$. Since $\gamma$ has order at least $k$, the powers $\gamma^r$ for $0 \leqslant r < k$ are all distinct. Also we know that $B$ is a *nonzero* polynomial of degree at most $s - 1$. Thus $B(\gamma^r) = 0$ for at most $s - 1$ values of $r$.

We have thus proved that the solution space has dimension at most $s - 1$. The claim about quadratic complexity and the structure of the matrix $M$ follows since the equations (12) of the linear system have a simple "lower-triangular" form. □

Combining Lemmas 3 and 6 and the decoding bound (9), we can conclude the following.

**Theorem 7.** *For the folded Reed-Solomon code $\mathrm{FRS}_q^{(m)}[n,k]$ of block length $N = n/m$ and rate $R = k/n$, the following holds for all integers $s$, $1 \leqslant s \leqslant m$. Given a received word $\mathbf{y} \in (\mathbb{F}_q^m)^N$, in $O((Nm \log q)^2)$ time, one can find a basis for a subspace of dimension at most $s-1$ that contains all message polynomials $f \in \mathbb{F}_q[X]$ of degree less than $k$ whose FRS encoding (1) differs from $\mathbf{y}$ in at most a fraction*

$$\frac{s}{s+1}\left(1 - \frac{mR}{m-s+1}\right)$$

*of the $N$ codeword positions.*

Note: When $s = m = 1$, the above just reduces to a unique decoding algorithm up to a fraction $(1-R)/2$ of errors.

**Comment on runtime and list size.** To get the actual list of close-by codewords, one can prune the solution subspace, which unfortunately may take $q^s$ time in the worst-case. This quantity is about $n^{O(1/\varepsilon)}$ for the parameter choices which achieve a list decoding radius of $1 - R - \varepsilon$. Theoretically, we are not able to improve the worst-case list size bound of $\approx n^{1/\varepsilon}$ in this regime. This motivates our results in Section 4 where we show that using a carefully chosen subset of all possible degree $k-1$ polynomials as messages, one can ensure that the list-size is much smaller while losing only a tiny amount in the rate.

Except for final step of pruning the subspace of candidate solutions, the decoding takes only quadratic time (and is perhaps even practical, as it just involves solving two structured linear systems). In practice, for example when errors occur randomly, the dimension of the output subspace will likely be very small, probably even 0 leading to a unique solution. If some side information about the true message $f$ is available that disambiguates the true message in the list [9], that might also be useful to speed up the pruning.

## 3 Some further comments about the proof method

We now make some salient remarks about the above linear-algebra based method to retrieve the space of polynomials $f$.

**Tightness of $q^{s-1}$ bound.** The upper bound of $q^{s-1}$ on the number of solutions $f$ to the Equation (10) cannot be improved in general. Indeed, let $A_0 = 0$, and $A_i$ for $1 \leqslant i \leqslant s$ be the coefficients of $Y^{i-1}$ in the polynomial $(Y-1)(Y-\gamma)\cdots(Y-\gamma^{s-2})$. Then for $0 \leqslant \ell \leqslant s-2$, we have

$$A_1 X^\ell + A_2(\gamma X)^\ell + \cdots + A_s(\gamma^{s-1} X)^\ell = X^\ell \cdot \left(A_1 + A_2 \gamma^\ell + A_3(\gamma^\ell)^2 + \cdots + A_s(\gamma^\ell)^{s-1}\right) = 0 \,.$$

By linearity, every polynomial $f \in \mathbb{F}_q[X]$ of degree at most $s-2$ satisfies (10). We should add that this does not lead to any non-trivial list-size lower bound for decoding folded RS codes, as we do not know if such a bad polynomial can occur as the output of the interpolation step, and moreover the pruning step could potentially reduce the size of the list further.

**Requirement on $\gamma$.** The argument in Lemma 6 only required that the order of $\gamma$ is at least $k$, and not that $\gamma$ is primitive. The polynomial $X^{q-1} - \gamma$ is irreducible if and only if $\gamma$ is primitive, and therefore the approach based on extension fields discussed in Section 2.1 requires $\gamma$ to be primitive. Usually in constructions of Reed-Solomon codes, one takes the block length $n \approx q$ and therefore the dimension $k$ is linear in $q$ (for constant rate codes). So this weakened requirement on $\gamma$ does not buy much flexibility in this case. However, if for some reason, one uses RS codes over much larger fields, then the new argument applies to a broader set of choices of evaluation points for the RS codes.

**Linear (instead of affine) space of solutions.** With a slight worsening of parameters, we can ensure that the space of solutions is in fact a *linear* space of dimension at most $s - 1$, instead of the affine space ensured by Lemma 6. The idea is to not use $A_0(X)$ in the interpolation (or rather set $A_0 = 0$), so that $Q(X, Y_1, Y_2, \ldots, Y_s) = A_1(X)Y_1 + A_2(X)Y_2 + \cdots + A_s(X)Y_s$. With the degree of each $A_i$ equal to $D$, this gives us $(D+1)s$ monomials in $Q$, and therefore condition (6) that guarantees the existence of a nonzero $Q$ meeting the interpolation requirements (7) now becomes $(D+1)s > N(m - s + 1)$. Thus one can take $D = \left\lfloor \frac{N(m-s+1)}{s} \right\rfloor$. The condition $t(m - s + 1) \geqslant D + k$ that enables successful list decoding is thus met when the agreement parameter $t$ satisfies $t \geqslant \frac{N}{s} + \frac{k}{m-s+1} = N\left(\frac{1}{s} + \frac{mR}{m-s+1}\right)$. This is slightly worse than (9), but still allows for decoding from agreement $(R + \varepsilon)N$ by setting $s \approx 1/\varepsilon$ and $m \approx 1/\varepsilon^2$.

**Hensel lifting.** An alternate approach (to root-finding over extension fields) for finding the low-degree solutions $f$ to the equation $Q(X, f(X), f(\gamma X), \ldots, f(\gamma^{s-1}X)) = 0$ is based on Hensel-lifting. Here the idea is to solve for $f \mod X^i$ for $i = 1, 2, \ldots$ in turn. For example, the constant term $f_0$ of $f(X)$ must satisfy $Q(0, f_0, f_0, \ldots, f_0) = 0$. If $Q(0, Y, Y, \ldots, Y)$ is a nonzero polynomial, then this will restrict the number of choices for $f_0$. For each such choice $f_0$, solving $Q(X, f(X), \ldots, f(\gamma^{s-1}X)) \mod X^2 = 0$ gives a polynomial equation for $f_1$, and so on. This approach is discussed in [1] and [4, Chap. 5]. It is mentioned that this algorithm is very fast experimentally and almost never explores too many candidate solutions. A similar approach was also considered in [16] for folded versions of algebraic-geometric codes. However, theoretically it has not been possible to derive any polynomial guarantees on the size of the list returned by this approach or its running time (the obvious issue is that in each step there may be more than one candidate value of $f_i$, leading to an exponential product bound on the runtime). Polynomial bounds in special cases (eg. when $s = 2$) are presented in [4], and obtaining such theoretical bounds is posed as an interesting challenge for future work. Our Lemma 6 provides an analysis of the Hensel-lifting approach when the interpolated polynomial is linear in the $Y_i$'s.

**Additive folding?** Let $p$ be a prime. Over $\mathbb{F}_p$, one can also consider *additive* folding schemes, where the value $f(a)$ is bundled together with $f(a+1), f(a+2), \ldots, f(a+m-1)$, in a construction similar to (1). The approach using extension fields can be used to show that the number of polynomials $f \in \mathbb{F}_p[X]$ of degree less than $p$ satisfying $Q(X, f(X), f(X+1), \ldots, f(X+s-1)) = 0$ for $Q \in \mathbb{F}_p[X, Y_1, Y_2, \ldots, Y_s]$ that is linear in the $Y_i$'s is at most $p^{s-1}$. This follows by going modulo the polynomial $X^p - X - 1$ which is irreducible over $\mathbb{F}_p$ and noting that $f(X+1) = f(X)^p \mod (X^p - X - 1)$.[5] Is there a linear-algebraic proof similar to Lemma 6 for this case? The map $f(X) \mapsto f(\gamma X)$ acts diagonally on the standard basis $\{1, X, \ldots, X^{k-1}\}$ for degree $k - 1$ polynomials, and this led to the nice structure for the linear system (10). The linear transformation $f(X) \mapsto f(X + 1)$ is not diagonalizable so the upper bound on the rank of the solution space may need a more careful

---

[5]The author first heard this argument for additive folding from Swastik Kopparty.

inspection of the structure of the system $A_0(X)+A_1(X)f(X)+A_2(X)f(X+1)+\cdots+A_s(X)f(X+s-1)=0$.

**Derivative codes.** Continuing the theme of the previous remark, when $\text{char}(\mathbb{F}_q) > k$, an analog of Lemma 6 for the differential equation

$$A_0(X) + A_1(X)f(X) + A_2(X)f'(X) + A_3(X)f''(X) + \cdots + A_s(X)f^{(s-1)}(X) = 0$$

is proved in [15] (here $f'(X)$ denotes the derivative of $f$ and $f^{(i)}(X)$ the $i$'th derivative of $f$). This is then used in [15] to show that *derivative codes* over fields of large characteristic can also achieve list decoding capacity. That is, they allow list decoding a fraction $1 - R - \varepsilon$ of errors with rate $R$, for a suitable choice of parameters. Independently, Bombieri and Kopparty [3] have given an algorithm for list decoding derivative codes up to a fraction $\approx 1 - R^{s/(s+1)}$ of errors using $s+1$-variate interpolation, matching the performance of the author and Rudra's algorithm for folded RS codes [13].

Derivative codes (or *univariate* multiplicity codes) are the variant of Reed-Solomon codes where the $i$'th codeword symbol consists of not only the value $f(a_i)$ at the $i$'th evaluation point, but also the values of its first $m-1$ derivatives (for some parameter $m \geqslant 1$). Over large characteristic, this is the same (up to some constant factors) as the residue of $f \mod (X - a_i)^m$. *Multivariate* versions of multiplicity codes were studied in the recent work of Kopparty, Saraf, and Yekhanin [18] where they were used to give a surprising construction of codes of rate $1 - \varepsilon$ locally decodable in $O(n^\gamma)$ time for any $\varepsilon, \gamma > 0$.

**Multiplicities, soft decoding, and list recovery.** For the linear interpolation of the form (4), using multiplicties in the interpolation stage, as in [14], only hurts the performance. This is because the degree of the $Y_i$'s cannot be increased to meet the needs of the larger number of interpolation conditions. Thus in order to get a good decoder than can handle *soft* information on reliabilities of various symbols [14, 17], one has to resort to the method behind the original algorithm in [13]. A weaker form of soft decoding is the problem of *list recovery*, where for each position $i$ of the code the input is a set $S_i$ of up to $\ell$ possible values, and the goal is to find all codewords whose $i$'th symbol belongs to $S_i$ for at least $t$ values of $i$. For this problem, a straightforward extension of the method of Section 2.2 gives an algorithm that works for agreement fraction $\tau = \frac{t}{N}$ satisfying

$$\tau > \frac{\ell}{s+1} + \frac{s}{s+1}\frac{mR}{m-s+1}.$$

The crucial point is that for any fixed $\ell$, by picking $s \approx \ell/\varepsilon$ and $m \approx \ell/\varepsilon^2$, we can list recover with agreement fraction $\tau = R + \varepsilon$ — the agreement fraction required does *not* degrade with increasing $\ell$. Such a list recovery guarantee is very useful in list decoding concatenated codes, for example to construct binary codes list-decodable up to the Zyablov radius, or codes list-decodable up to radius $1 - R - \varepsilon$ over alphabets of fixed size independent of $n$; see [13, Sect. V].

## 4 Improving list size via pseudorandom subspace-evasive subsets

Based on Theorem 7, in this section we pursue one possible approach to improve the provable worst-case list size bound for list decoding up to a fraction $1 - R - \varepsilon$ of errors. Instead of allowing all polynomials $f_0 + f_1 X + \cdots + f_{k-1} X^{k-1}$ of degree less than $k$ as messages, the idea is to restrict the coefficient vector $(f_0, f_1, \ldots, f_{k-1})$ to belong to some special subset $\mathcal{V} \subseteq \mathbb{F}_q^k$, satisfying the following two conflicting demands:

**Largeness:** The set $\mathcal{V}$ must be large, say $|\mathcal{V}| \geqslant q^{(1-\varepsilon)k}$, so that the rate is reduced by at most a $(1-\varepsilon)$ factor.

**Low intersection with subspaces:** For every subspace $S \subset \mathbb{F}_q^k$ of dimension $s$, $|S \cap \mathcal{V}| \leqslant L$.
(Let us call this property of $\mathcal{V}$ as $(s, L)$-*subspace-evasive* for easy reference. The field $\mathbb{F}_q$ and ambient dimension $k$ will be fixed in our discussion.)

Using such a set $\mathcal{V}$ will ensure that after pruning the affine subspace output by the algorithm of Theorem 7, the number of codewords will be at most $L$. (Note that an affine subspace of dimension $s-1$ is contained in a subspace of dimension $s$.) Thus the list size will go down from $q^{s-1}$ to $L$.

Subspace-evasive subsets were used in [21] to construct bipartite Ramsey graphs, and in fact we borrowed the term *evasive* from that work. In their work, the underlying field was $\mathbb{F}_2$ and the subsets had to be evasive for dimension $s \approx k/2$. Our interest is in a different (and hopefully easier?) regime — we can work over large fields, and are interested in evasiveness w.r.t. $s$-dimensional subspaces for constant $s$.

Subspace-evasive subsets are also connected to certain well-studied objects called *affine extractors*; see the discussion at the end of this section.

A random large subset of $\mathbb{F}_q^k$ meets the low subspace intersection requirement very well, as shown below. The argument is straightforward; a similar bound appears in [5] in the geometric context of point-subspace incidences.

**Lemma 8.** *Let $\mathcal{W}$ be a random subset of $\mathbb{F}_q^k$ chosen by including each $x \in \mathbb{F}_q^k$ in $\mathcal{W}$ with probability $q^{-s-\alpha}$ for some $\alpha > 0$. Then with probability at least $1 - q^{-\Omega(k)}$, $\mathcal{W}$ satisfies both the following conditions: (i) $|\mathcal{W}| \geqslant q^{k-s-\alpha}/2$, and (ii) $\mathcal{W}$ is $(s, 2sk/\alpha)$-subspace-evasive.*

*Proof.* The first part follows by a standard Chernoff bound calculation. For the second part, fix a subspace $S \subseteq \mathbb{F}_q^k$ of dimension $s$, and a subset $T \subseteq S$ of size $t = \lceil 2ks/\alpha \rceil$. The probability that $\mathcal{W} \supseteq T$ equals $q^{-(s+\alpha)t}$. By a union bound over the at most $q^{ks}$ choices for the $s$-dimensional subspace $S$, and the at most $q^{st}$ choices of $t$-element subsets $T$ of $S$, we get that the probability that $\mathcal{W}$ is not $(s, t-1)$-subspace-evasive is at most $q^{ks+st} \cdot q^{-(s+\alpha)t} \leqslant q^{-ks}$ since $t \geqslant 2ks/\alpha$. □

Picking $\alpha \approx \varepsilon k$, the above guarantees the existence of subsets $\mathcal{W}$ of $\mathbb{F}_q^k$ of size $q^{(1-\varepsilon)k}$ which are $(s, O(s/\varepsilon))$-subspace-evasive. Restricting the coefficient vector $(f_0, f_1, \ldots, f_{k-1})$ of the message polynomial to belong to such a subset will guarantee a list-size upper bound of $O(s/\varepsilon)$ in Theorem 7. This list-size bound is a *constant independent of $n$*, and for the choice $s \approx 1/\varepsilon$ which enables list decoding a fraction $1 - R - \varepsilon$ of errors, it is $O(1/\varepsilon^2)$. This is quite close to the bound of $O(1/\varepsilon)$ achieved by random codes [11].

Unfortunately, an explicit construction of subspace-evasive subsets with anywhere close to the trade-off guaranteed by the probabilistic construction of Lemma 8 is not known. This appears to be a challenging and extremely interesting question. One natural choice for such a subset would be some *variety* $\mathcal{V} \subseteq \mathbb{F}_q^k$ defined by a collection of polynomial equations, i.e., $\mathcal{V} = \{\mathbf{a} \in \mathbb{F}_q^k \mid g_1(\mathbf{a}) = g_2(\mathbf{a}) = \cdots = g_l(\mathbf{a}) = 0\}$ for some polynomials $g_1, g_2, \ldots, g_l \in \mathbb{F}_q[Z_1, Z_2, \ldots, Z_k]$. Indeed for $s = 1$ and $s = k - 1$, varieties in $\mathbb{F}_q^k$ (the modular moment surface and modular moment curve) with low intersection with $s$-dimensional affine subspaces are known [5].

**Connection to affine extractors.** The problem of constructing subspace-evasive subsets is related to the well-studied problem of constructing *affine extractors*. An affine extractor is an $M$-coloring of $\mathbb{F}_q^k$ with the property that every $s$-dimensional affine subspace of $\mathbb{F}_q^k$ has between $(1/M - \delta)q^s$ and $(1/M + \delta)q^s$ elements belonging to each of the $M$ color classes. Here $\delta$ is the error and $\log_2 M$ is the number of output bits of the extractor. If we had an affine extractor with a large number of outputs (say $M \geq q^{(1-\varepsilon)s}$ for arbitrarily small constants $\varepsilon > 0$) and very small error ($\delta \leq O(1/M)$, in other words a small *relative* error instead of an additive error), then the subset corresponding to a single color class will be subspace-evasive.

Known explicit constructions of affine extractors fall short of meeting these requirements. The constructions in the literature either require large dimensions $s$ (and therefore are not applicable in our setting of $s = O(1)$), or have too large an error to be useful for us. For instance, the extractor of Gabizon and Raz [7], which works over large fields and any $s \geq 1$ (both aspects being perfect for us), has an error $\delta \approx 1/\sqrt{q}$, due to the application of the Weil bounds on character sums. On the other hand, an extractor satisfies a stronger property than what is needed in a subspace-evasive subset, so we hope that good explicit constructions of subspace-evasive subsets will be easier to obtain.

### 4.1 Pseudorandom construction of subspace-evasive subsets

The construction of Lemma 8 takes exponential time and produces a random unstructured set that takes exponential space to store. In this section, we show that a subset with similar guarantees can be constructed in probabilistic polynomial time, producing a polynomial size representation of the constructed subspace-evasive set. The idea is to note that the probabilistic argument to argue about $(s,t)$-subspace-evasiveness only needed $t$-wise independence and not complete independence of different elements of $\mathbb{F}_q^k$ landing in the random subset $\mathcal{W}$.

Fix an arbitrary basis $1, \beta, \beta^2, \ldots, \beta^{k-1}$ of $\mathbb{F}_q^k$ over $\mathbb{F}_q$. Also denote $\mathbb{K} = \mathbb{F}_{q^k}$. For a polynomial $P \in \mathbb{K}[X]$ and an integer $r$ ($1 \leq r \leq k$), define the subset $\mathbb{S}(P, r) \subseteq \mathbb{F}_q^k$ as follows:

$$\mathbb{S}(P,r) = \{(a_0, a_1, \ldots, a_{k-1}) \in \mathbb{F}_q^k \mid P(a_0 + a_1\beta + a_2\beta + \cdots + a_{k-1}\beta^{k-1}) \in \mathbb{F}_q\text{-span}(1, \beta, \cdots, \beta^{r-1})\}.$$

**Lemma 9.** *Let $q$ be a prime power, $k \geq 1$ an integer, and denote $\mathbb{K} = \mathbb{F}_{q^k}$. Let $\zeta \in (0,1)$ and $s$ be an integer satisfying $1 \leq s \leq \zeta k/2$. Let $P \in \mathbb{K}[X]$ be a random polynomial of degree $t$ and define $\mathcal{V} = \mathbb{S}(P, (1-\zeta)k)$. Then, provided $t \geq \Omega(s/\zeta)$, with probability at least $1 - q^{-\Omega(k)}$ over the choice of $P$, $\mathcal{V}$ is a $(s,t)$-subspace-evasive subset of $\mathbb{F}_q^k$ of size at least $q^{(1-\zeta)k}/2$.*

*Proof.* For each $\mathbf{x} \in \mathbb{F}_q^k$, note that $\mathbf{x} \in \mathbb{S}(P, r)$ with probability $q^{-\zeta k}$. Further, since the values of $P$ at any $t$ distinct points in $\mathbb{K}$ are independent, the events $\mathbf{x} \in \mathbb{S}(P, r)$ for various $\mathbf{x} \in \mathbb{F}_q^k$ are $t$-wise independent. The argument in Lemma 8 only relied on the $t$-wise independence of these events, and therefore one can conclude that $\mathcal{V} = \mathbb{S}(P, r)$ is $(s, O(s/\zeta))$-subspace-evasive with probability at least $1 - q^{-\Omega(ks)}$.

The expected size of $\mathcal{V}$ is $\mathbb{E}[\mathcal{V}] = q^{(1-\zeta)k}$. Since the events $\mathbf{x} \in \mathcal{V}$ are pairwise independent, by Chebyshev's inequality, $\Pr[|\mathcal{V}| < \mathbb{E}[\mathcal{V}]/2] < 4/\mathbb{E}[\mathcal{V}]$. Hence $|\mathcal{V}| \geq q^{(1-\zeta)k}/2$ except with probability at most $q^{-\Omega(k)}$. □

Note that the set $\mathbb{S}(P,r)$ has a compact representation, and given $P$, membership in $\mathbb{S}(P,r)$ can be checked efficiently. In fact, it is easy to see that $\mathbb{S}(P,r)$ is a variety in $\mathbb{F}_q^k$. Indeed, $P(a_0 + a_1\beta + \cdots + a_{k-1}\beta^{k-1})$ can be expanded out as $p_0(a_0, a_1, \ldots, a_{k-1}) + p_1(a_0, \ldots, a_{k-1})\beta + \cdots + p_{k-1}(a_0, \ldots, a_{k-1})\beta^{k-1}$ for polynomials $p_0, p_1, \ldots, p_{k-1} \in \mathbb{F}_q[Z_1, Z_2, \ldots, Z_k]$, and therefore

$$\mathbb{S}(P,r) = \{\mathbf{a} = (a_0, a_1, \ldots, a_{k-1}) \in \mathbb{F}_q^k \mid p_r(\mathbf{a}) = p_{r+1}(\mathbf{a}) = \cdots = p_{k-1}(\mathbf{a}) = 0\}. \tag{13}$$

Combining this with Theorem 7, we can conclude the following.

**Theorem 10.** *For any $\zeta > 0$, there is a Monte Carlo construction of a subcode $C$ of $\mathrm{FRS}_q^{(m)}[n,k]$, consisting of encodings of polynomials whose coefficients belong to a variety $\mathcal{V} \subset \mathbb{F}_q^k$, such that with high probability $C$ has rate at least $(1-\zeta)k/n$ and can be list decoded from a fraction $\frac{s}{s+1}\left(1 - \frac{mR}{m-s+1}\right)$ for any $1 \leqslant s \leqslant m$ in $q^{O(s)}$ time with an output list size of at most $O(s/\zeta)$.*

*In particular, picking $\zeta = \Theta(\varepsilon)$, $s = \Theta(1/\varepsilon)$ and $m = \Theta(1/\varepsilon^2)$, the construction yields codes of rate $R$ which can be list decoded from a fraction $1 - R - \varepsilon$ of errors in polynomial time, with at most $O(1/\varepsilon^2)$ codewords output in the list.*

**Encoding complexity.** In the above construction, the code can be succinctly stored and membership in the code efficiently tested. However, we do not know a way to output the $i$'th codeword in the code (i.e., to perform encoding) in polynomial time. We now show that efficient encoding can also be achieved if we settle for a list size of $O(k)$ (which is still much better than the $q^{\Omega(1/\varepsilon)}$ bound).

The idea is to apply Lemma 9 with the parameter choice $\zeta = 2s/k$ and $t = O(k)$, and taking $\mathcal{V} = \mathbb{S}(P, k-2s)$ for a random degree $t$ polynomial $P$. Now with very high probability over the choice of $P$, standard tail inequalities for $t$-wise independent random variables (eg. [2]) imply that for every choice of $f_0, f_1, \ldots, f_{k-3s-1}$, there are at least $q^s/2$ elements $(a_0, a_1, \ldots, a_{k-1}) \in \mathbb{S}(P,r)$ such that $a_i = f_i$ for $0 \leqslant i < k - 3s$. In particular such a $k$-tuple can be found in $q^{O(s)}$ time by searching over all possible values of $(a_{k-3s}, \ldots, a_{k-1})$. We can use an arbitrary such tuple $(a'_{k-3s}, \ldots, a'_{k-1})$ (say the lexicographically smallest) as the $3s$ highest degree coefficients and encode the message $(f_0, f_1, \ldots, f_{k-3s-1}) \in \mathbb{F}_q^{k-3s}$ by the folded RS encoding (1) of the polynomial $f_0 + f_1 X + \cdots + f_{k-3s-1}X^{k-3s-1} + a'_{k-3s}X^{k-3s} + \cdots + a'_{k-1}X^{k-1}$. Note that we only purge $3s$ symbols from $\mathbb{F}_q$ in the messages so the rate is $R - o(1)$. The list decoder can simply discard the top (highest degree) $3s$ coefficients of any recovered polynomial to find the actual message tuple.

One obvious open question raised by the above is to construct the claimed variety (even with a somewhat worse list size guarantee) explicitly. This would make the code explicit, and if the variety is sufficiently well-structured, also imply a nice encoding function. Even more exciting would be to construct a subspace-evasive subset for which the intersection with an $s$-dimensional subspace can be computed efficiently, in time polynomial in the size of the intersection. This would avoid the need for the $q^s$ runtime bottleneck arising from exhaustively checking all candidates in the subspace for membership in the variety.

One point worth noting is that the degree of each of the polynomials $p_i \in \mathbb{F}_q[Z_1, Z_2, \ldots, Z_k]$ defining the variety (13) is $\Omega(s/\zeta)$ and there are $\zeta k$ of them, so bounding the size of the variety by the product of the degrees via Bezout's theorem would lead to uselessly large bounds. Even the *existence* of a variety cut out by say $O(s)$ polynomials each with degree at most $O(s)$ that is $(s, s^{O(s)})$-subspace-evasive does not appear to be known.

## Acknowledgments

I am grateful to Salil Vadhan for telling me about the elegant degree 1 interpolation method for list decoding folded RS codes. I thank Carol Wang for useful discussions and Atri Rudra for valuable comments on the write-up. I thank Noga Alon, Swastik Kopparty, Po-Shen Loh, and David Zuckerman for their input on the literature on subspace-evasive sets, and Ran Raz for discussions about their construction. Thanks to Swastik for pointing me to [21], and to Po for pointers to work on related concepts in geometric settings [6].